\documentstyle[twocolumn,prl,aps]{revtex}
\input epsf
\begin{document} %\draft
%\twocolumn
\title{
Unlocking Transition for Modulated Surfaces and Quantum Hall Stripes 
}
\author{
H.A.~Fertig
}
\address{
Department of Physics and Astronomy, University of Kentucky,
Lexington, KY 40506-0055
}
\address{\mbox{ }}
\address{\parbox{14.5cm}{\rm \mbox{ }\mbox{ }
We develop a sine-Gordon model of layered systems of
two-dimensional modulated surfaces and one dimensional stripes,
and demonstrate that these systems can undergo a
Kosterlitz-Thouless transition
in which the modulations unlock as a result of 
thermal or quantum fluctuations, respectively.  The unlocked phase
is interpreted as an anisotropic crystal in which soliton-antisoliton
pairs proliferate.  The properties of such a state 
for modulated stripes in quantum Hall systems and its possible
relevance to recent anomalies in transport data are discussed.  
}}
\address{\mbox{ }}
\address{\mbox{ }}
%\end{abstract}

\maketitle

\begin{figure}
 \vbox to 6.0cm {\vss\hbox to 10cm
 {\hss\
   {\includegraphics{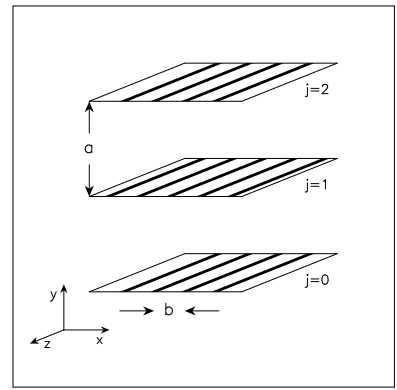}
   }
  \hss}
 }
\vspace{2mm}
\caption{Example of a modulated sheet system.  Shaded areas indicate
regions where the two-dimensional sheet density is larger than the
average.  For repulsive intersheet interactions,
at low temperatures one expects the modulations in neighboring
sheets to be shifted with respect to one
another, forming an anisotropic crystal.}
\label{sheets}
\end{figure}

Systems which may be modeled as interacting elastic surfaces occur in many 
different contexts in nature, including liquid crystals, domain walls
in magnets\cite{lub}, electron gases in semiconductor 
superlattices\cite{dassarma}, layered superconductors\cite{horovitz}, and 
biophysical systems\cite{lau}.  In many situations at a mean-field
level the surfaces themselves may contain spatially periodic structure  
as illustrated in Fig. \ref{sheets}.  
At very low temperatures, such modulations on different sheets
should be strongly correlated so that the system forms a type
of three dimensional crystal.
An interesting
question one may ask is: if the modulations of
the surfaces are weakly coupled, can thermal fluctuations
cause the modulations on different
sheets to become
uncorrelated  -- i.e., unlocked -- even
if thermal fluctuations do not
disorder the modulations
within a given surface\cite{lub2}? 
In a precise analogy\cite{sondhi}, one can also ask whether a series of modulated elastic
{\it lines} will remain locked at zero temperature when quantum fluctuations
are taken into account.  It is this latter problem that motivates
this work and will be our principle focus.

The modulated elastic line problem is motivated by the theoretical discovery
of striped electronic phases in quantum Hall systems in high Landau 
levels\cite{koulakov,moessner,prange}.  Hartree-Fock studies of such
striped phases at zero temperature\cite{fertig,macdonald} indicate 
that they are 
generically unstable
to the formation of modulations along the stripes, 
so that at the mean-field level the stripe state is essentially
an electron (Wigner) crystal\cite{ferrev}, albeit a highly anisotropic
one.  Recent interest in this system has grown due to the 
discovery\cite{lilly,du}
of strong anisotropies in the transport properties of  
high quality two-dimensional electron 
systems in perpendicular magnetic fields, between
quantum Hall plateaus corresponding to 
filling factors $\nu > 4$.  ($\nu \equiv 2\pi \rho_0 l_0^2$,
$\rho_0$ is the electron density, $l_0^2 = \hbar c/eB$
and $B$ is the magnetic field.)
In these experiments, dc transport data at very low
temperature exhibit a dissipative
linear conductivity that is much greater in one
direction than the other (i.e., $\sigma_{xx} \gg \sigma_{yy}$
for some direction $x$\cite{dir}).
As a function of 
filling factor, $\sigma_{xx}$ exhibits a strong peak
around $\nu_x=1/2$, where $\nu_x$ is the fractional
part of the filling factor, while $\sigma_{yy}$ has
a {\it minimum} around $\nu_x=1/2$.  The system also
exhibits unusual non-linear transport: for large applied
currents in the direction of high conductivity, the
the dissipation is greater than what would be expected
either from linear response or from a simple heating model\cite{lilly}.
We will argue below that these properties may be understood
in the framework of a modulated stripe phase, unlocked by
quantum fluctuations\cite{fradkin}.

We begin by defining a simple model 
of sheets with modulations present in them (see Fig. \ref{sheets}).  The
Hamiltonian may be written as $H=H_0+H_{\lambda}$, with
\begin{eqnarray}
H_0&=&{1 \over 2}\sum_j\int{dxdz} \bigl\lbrace 
\kappa_x |{\partial_x}u^x_j|^2 + 
\kappa_z |{\partial_z}u^x_j|^2 \bigr\rbrace \nonumber \\
&&+{1 \over 2} U \sum_j \int{dxdz}  \bigl[u_j^y - u_{j-1}^y]^2,
\label{h0membrane}
\end{eqnarray}
\begin{equation}
H_{\lambda} = -\lambda \sum_j \int dxdz 
\cos\Bigl[{{2\pi} \over b} \bigl(u_j^x(x,z) - u_{j-1}^x(x,z)\bigr) \Bigr].
\label{hlambda}
\end{equation}
The fields $u_j^{\mu}(x,z)$ represent displacements
of the $j~th$ layer in the direction $\mu$ and are the
dynamical variables for this problem, $\kappa_{\mu}$ are elastic
constants for a given layer, and $U$ represents the energy 
scale of a harmonically-approximated repulsion between layers.
In $H_{\lambda}$, the $b$ inside the cosine represents the period
of the layer modulations.
Since $b$ is the natural
length scale in this problem, we will set it to 1 and adopt
it as our unit of length\cite{dislocation}.

To develop some intuition as to what can happen in this model, it is
convenient to ``freeze'' all the sheets except one; i.e., set
$u^{\mu}_j=0$ except for $j=0$.  If one computes the partition function
$Z=\int {\cal D}\vec{u}_0 e^{-H}$ it is easy to see that this separates into a
product of two functional integrals, one for each direction of
displacement.  Of the two integrals, only the one over $u^x_0$
has a non-trivial structure; it is precisely the partition function for
the two-dimensional sine-Gordon model.  
For small $\lambda$
it is well-known that this model has
a Kosterlitz-Thouless phase transition\cite{lub}.
For large values of $\kappa_x$, the $j=0$ sheet will be locked into
what we have approximated as the periodic potential of the other
(for the moment) static sheets.  This state has a set of interesting
excitations in the form of solitons.  
At any temperature one should expect to see occasional soliton-antisoliton
pairs in the system, but such
soliton pairs should always recombine.  For the dynamical ($j=0$) layer,
such an excitation represents a finite size patch that has slipped by
one period of the modulation ($b$) relative to the other layers.
As $\kappa_x$ is decreased, the soliton-antisoliton pairs become
more tenuously bound, and the sizes of the patches become increasingly
large.  At a critical value of $\kappa_x$, the solitons fully unbind,
leading to arbitrarily large slips of the dynamical surface, and
hence its unlocking from the modulations of the others.

In the 2+1 dimensional quantum problem of modulated lines, 
the soliton-antisoliton
pairs have a simple interpretation: they are vacancy-interstitial pairs
forming on a given line.  Vacancies and interstitials in a Wigner crystal
can carry current\cite{ferrev}; thus, if quantum fluctuations
populate the ground state with unbound pairs, we expect the system to
behave as a metal rather than an insulator.  This will be discussed
in more detail below.

To properly treat the unlocking transition, we need to include the
dynamics of all the layers.  A convenient way to approach
this is to treat $H_{\lambda}$ as a perturbation, and compute the change
in the effective elastic constants $\kappa_x$ and $\kappa_z$ due
to its presence.  Following a standard procedure\cite{lub}, we
define the effective stiffnesses of the system by imposing a gradient
in the displacement field: $u^{x}_j(x,z)=u^{x(0)}_j(x,z) + 
\sum_{q_y} e^{iq_yja}\vec{v}_{q_y}\cdot \vec{x}$ where $\vec{x}=(x,z)$, 
$\vec{v}_{q_y}=\vec{v}^{*}_{-q_y}$, 
$a$ is the distance between layers, 
and the field
$u^{x(0)}_j(\vec{x})$ must vanish at the system boundaries.
The free energy of the system should then take the form
$F(v)-F(0)={\Omega \over {2}} \sum_{q_y} \sum_{\mu,\nu}
\kappa^R_{\mu,\nu}(q_y) v^{\mu}_{q_y} v^{\nu}_{-q_y}$ where 
$\kappa^R_{\mu,\nu}(q_y)$ is an effective elasticity tensor,
$\Omega$ is the
area of a sheet in the $x,z$ plane,
and $\mu,~\nu=x,~z$.
Note the $q_y$ dependence tells us that the introduction of $H_{\lambda}$
couples displacements in different layers together, even if they
are not fully locked.

Using the above expression for the free energy, the renormalized
coupling constants may be computed\cite{elsewhere} to $O(\lambda^2)$.
To this order
the resulting elastic constants are diagonal in their
indices ($\kappa_{\mu,\nu}^R = \kappa_{\mu}^R \delta_{\mu,\nu}$),
and have the form
\begin{equation}
\kappa_{\mu}^R(q_y) = \kappa_{\mu}
+{{(2\pi\lambda)^2} \over {4 \Omega}} \gamma(q_y)
W_{\mu} \int_{a_c}^{\infty} dr r^3 \Bigl( {r \over {a_c}} \Bigr)^{-x_{\kappa}}. 
\label{kppar2}
\end{equation}
In the above expression, $\gamma(q_y)=2-2cos(q_ya)$, 
$(W_{x},W_z)=\int_0^{2\pi} d\theta (\cos^2\theta,\sin^2\theta)
e^{-f_{\theta}}$, with $f_{\theta} = \pi a  \int 
{{dq_y} \over {2\pi}} \gamma(q_y) ln[{{\kappa_z} \over {\kappa_x}}\cos^2\theta
+\sin^2\theta]/(\kappa_x \kappa_z)^{1/2}$, $a_c$ is an ultraviolet
cutoff of order $b$,
and
$$
x_{\kappa} = a \int dq_y {{\gamma(q_y)} \over
{(\kappa_z \kappa_x)^{1/2}}}.
$$
The quantity $x_{\kappa}$ plays a central role in this model.
One may see that the renormalized elasticity constants diverge
in perturbation theory if 
$x_{\kappa} \le 4$.  We take this divergence to signal the
onset of the locking transition between the layers.
The form of Eq. \ref{kppar2} is typical for a 
system that undergoes a Kosterlitz-Thouless
transition\cite{lub}.  

\begin{figure}
 \vbox to 6.0cm {\vss\hbox to 8cm
 {\hss\
   {\includegraphics{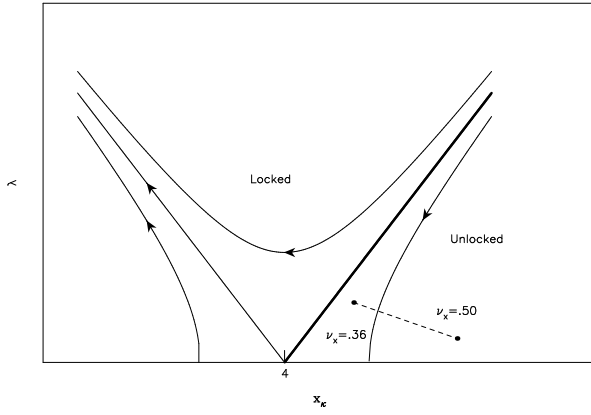}
   }
  \hss}
 }
\vspace{2mm}
\caption{Phase boundary and renormalization group flows for unlocking
transition.  Bold line denotes phase boundary.
Dotted line illustrates possible initial conditions
relevant to the striped phase, discussed in the text.  
It is assumed (see Ref. 5)
that below $\nu_x \approx 0.36$ the system is in an insulating ``bubble
phase'' 
%\cite{koulakov}% 
and enters the unlocked stripe phase
via a first order transition.  Filling factors above $1/2$ are
related to those below by particle-hole symmetry.}
\label{flows}
\end{figure}

To get an approximation of the phase diagram for this system, 
it is convenient to
expand the elastic constants in a complete set of states, 
$\kappa_{\mu}(q_y) = \sum_n \kappa_{\mu}^{(n)} \gamma_n (q_y)$,
setting $\gamma_0 = [2-2\cos(q_ya)]/\sqrt{6}$.  One can then
see that to order $\lambda^2$ only $\kappa_{\mu}^{(0)}$ is actually
renormalized by the interlayer coupling.  To order $\lambda^2$ the
scaling relations are 
\begin{eqnarray}
{{dx_{\kappa}} \over {d\ell}} &=& -\lambda^2 C(\ell) \nonumber \\
{{d\lambda} \over {d \ell}} &=& {1 \over 2} [4-x_{\kappa}]\lambda
\label{scaling}
\end{eqnarray}
with $C(\ell)= \sqrt{6} a \int dq_y \gamma_0(q_y){1 \over {(\kappa_x
\kappa_z)^{1/2}}}\sum_{\mu} {{W_{\mu}} \over {\kappa_{\mu}}}$.  Since
$C(\ell)$ behaves smoothly in the vicinity of the critical point, 
we can set $C(\ell) \rightarrow C(0)$ and only incur errors of order
$\lambda^3$.  With this substitution, one can derive the
phase boundary and renormalization group flows in the 
$(x_{\kappa},\lambda)$ plane.
The result is illustrated in Fig. \ref{flows}. 
It is interesting to note that similar phase diagrams
have been obtained in studies of 
Josephson coupled, layered superconductors\cite{horovitz}.

We now turn to the application of this model to the striped phase
in the quantum Hall system.  Hartree-Fock studies of striped
phases\cite{koulakov,moessner} in high Landau levels 
indicate that, within mean-field theory,
they are
unstable to the formation of weak modulations
within each stripe\cite{fertig,macdonald}.  
The modulation period that is
favored turns out to be precisely what is needed so the resulting
anisotropic two-dimensional crystal has one
electron per unit cell.  
Qualitatively, the modulations appear to get weaker as the
fractional part of the filling factor, $\nu_x$, approaches
1/2; however, they never completely vanish, and at $\nu_x = 1/2$
the mean-field state has broken particle-hole symmetry.

To model fluctuations around this mean-field state, we consider 
each stripe as an elastic line\cite{com},
with the modulations coupled by the Hamiltonian $H_{\lambda}$ above.
In the absence of modulations, our model potential energy is 
\begin{eqnarray}
V_0 = {1 \over 2} U \sum_j \int dx \bigl[u_j^y(x)-
u_{j-1}^y(x)\bigr]^2 \nonumber \\
+ {1 \over 2} \sum_j \int dx 
\kappa_x \Bigl( {{d u_j^x(x)} \over {dx}} \Bigr)^2.
\label{newH0}
\end{eqnarray}
This model Hamiltonian is most appropriate for electrons interacting
via short-range interactions; the effects of long-range interactions
will be discussed elsewhere\cite{elsewhere}.
In the quantum Hall regime, we project the dynamics of the
stripes into a single Landau level
by making the replacement\cite{kubo} $u^y_j(x) \rightarrow {{l_0^2} \over i}
{{\partial} \over {\partial u^x_j(x)}}$.
In the standard fashion\cite{sondhi}, one can now compute most quantities 
of interest
from the generating functional $Z=\int {\cal D} ue^{-S}$, with
$S=H_0+H_{\lambda}$, and\cite{elsewhere}
\begin{eqnarray}
H_0 = 
{1 \over {2}} \int dxd\tau 
%\Biggl\lbrace
\sum_{q_y} \Biggl[ &\kappa_z(q_y)& \Biggl| {{du(x,q_y,\tau)} \over {d\tau}} 
\Biggr|^2
\nonumber \\
&+&\kappa_x  \Biggl| {{du(x,q_y,\tau)} \over {dx}} \Biggr|^2 
\Biggr] 
%\Biggr\rbrace, \nonumber \\
\end{eqnarray}
where $u(x,q_y,\tau) \equiv {1 \over {\sqrt{N_c}}} \sum_j e^{-iq_yja} 
u_j^x(x,\tau)$,
$N_c$ is the number of stripes, and $1/\kappa_z(q_y)=U\gamma(q_y)l_0^4$.
$H_{\lambda}$ is formally identical to Eq.
\ref{hlambda}.  Relabeling $\tau \rightarrow z$, except for the
$q_y$ dependence in the ``bare'' value of $\kappa_z$,
the system is formally identical to the 
one studied above.  The analysis follows through
essentially without change.   

The structure of the phase diagram in Fig. \ref{flows} suggests that
if the stiffnesses of the system $\kappa_{\mu}$ are large enough, or
more precisely if $x_{\kappa}$ is small enough, then the system will
remain in a locked phase even for arbitrarily small amplitude
modulations $\lambda$. 
Writing $x_{\kappa}$ in terms of the original parameters $\kappa_x$
and $U$, one finds that  
$\Bigl[ {{\kappa_x b^4} \over {Ul_0^4}} \Bigr]^{1/2} < 16/3$
to produce an unlocked, free soliton phase.  
For arbitrarily weak interactions one can estimate
the value of this parameter\cite{elsewhere};  
the result is
$\Bigl[ {{\kappa_x b^4} \over {Ul_0^4}} \Bigr]^{1/2} 
= \sqrt{2}\pi < 16/3$.  Thus, for weak interactions, we expect
the modulations to unlock.  We note that it is not immediately obvious 
what happens
as repulsive interactions are turned up: these increase both 
$\kappa_x$ and $U$, and whether their ratio is increased or decreased
depends upon the microscopic details of the interaction.  Studies
of this are currently underway.

In direct analogy with the unpinned phase of the
sine-Gordon equation, the unlocked
phase may be thought of as a highly anisotropic Wigner crystal state,
in which free solitons -- vacancies and interstitials of the Wigner
crystal -- are included in the ground state due to quantum fluctuations.
If the disorder is weak enough, or the temperature high enough, so that
localization of the solitons may be ignored\cite{com3}, these
solitons can carry current through the system.
The properties of the unlocked phase of this anisotropic
Wigner crystal turn out 
to be consistent with many aspects
of the experimental data.  (1) Charge transport through the
system would be highly anisotropic.
Clearly, the solitons are far more
mobile along the stripes than across them.  Transport across the
stripes requires tunneling of the solitons, whose
amplitude should be small since the stripe modulations
in the mean-field state are weak\cite{com4}.  (2) In experiment,
a peak in the longitudinal conductivity is observed for some
direction of transport (e.g., in $\sigma_{xx}$) around $\nu_x=1/2$.
If one assumes that
the parameters $x_{\kappa}$ and $\lambda$ follow a trajectory as
a function of $\nu_x$ such as that 
shown in Fig. \ref{flows}, then one moves more
deeply into the unlocked phase as $\nu_x \rightarrow 1/2$.  
The number of carriers -- solitons -- then would {\it increase} 
as one approaches $1/2$. 
Due to an approximate particle-hole symmetry in the
microscopic Hamiltonian\cite{fertig}, this leads to a peak in
$\sigma_{xx}$ if the stripes lie along the $x$-direction.
A tantalizing possibility for this system is that it can undergo a 
continuous quantum phase transition as a function of $\nu_x$
if the microscopic parameters $x_{\kappa}$ and $\lambda$ pass through
the phase boundary (bold line in Fig. \ref{flows}) as a function of $\nu_x$.  
We note, however, that mean-field calculations\cite{koulakov}  
support a scenario in which the specific sample
studied in Ref. \onlinecite{lilly}
has a first order transition directly into 
an (unlocked) striped phase.  
(3) In experiment, a dip is observed in $\sigma_{yy}$
around $\nu_{x}=1/2$.  In the unlocked phase,
linear transport perpendicular
to the stripes depends both on the density of solitons and the
amplitude for soliton tunneling across stripes.  Since the modulations
of the stripes become more well-developed as one moves away from
$\nu_x=1/2$ \cite{fertig}, this amplitude should be
an increasing function of $|\nu_x-1/2|$ \cite{com4}.  If the amplitude
changes fast enough, one would expect to see a dip in $\sigma_{yy}$.  
A microscopic study of the
soliton tunneling amplitude is currently underway.  
(4) Large currents (in the same
direction as that of a measurement of $\sigma_{xx}$) 
lead to anomalously
large dissipation in experiment.  For the unlocked
stripe phase, such a large current would be accompanied by
a large electric field parallel to the stripes\cite{elsewhere}.
Such a field would generate
soliton-antisoliton pairs, leading to the
enhanced dissipation. 

Finally, we comment briefly on other related 
models currently in the literature.
In Refs. \onlinecite{koulakov,moessner} striped phases were predicted
but the instability to modulations in mean-field theory
were not noted.  In Ref. \onlinecite{fradkin} the instability
is pointed out, shape fluctuations are proposed as a mechanism
to restore the stripe phase, and any further effects of modulations
on the stripes are not considered.  
In all these models, tunneling between
stripes at zero temperature is not possible due to symmetry
considerations, leaving open the question of why $\sigma_{yy}$ is
non-vanishing in experiment.  Presumably, disorder lifts the symmetry
and allows interstripe transport in such models, but to
explain the experiments one
must adopt a disorder model that allows interstripe tunneling
but not localization effects
at the lowest experimentally available temperatures.
By contrast, in the model discussed here the energy scale 
for tunneling is set by
electron-electron interactions, 
so for weak disorder it is natural for
the system to behave metallically over a range
of temperatures.  
We note also that an interesting possible route to interstripe
tunneling, discussed in Ref. \onlinecite{fradkin}, is via
dislocations if they proliferate (leading to a nematic phase)
due to either thermal or quantum fluctuations.  Whether the
fluctuations in experiment are strong enough to melt the
stripe phase in this way is currently unknown; but if present
they should lead to interstripe transport in parallel to that
discussed in this work.
Finally, when described
in terms of free vacancies and interstitials the
present model admits a simple explanation for the nonlinear
transport properties seen in experiment.  It is unclear
at present how unmodulated stripes might yield
such behavior.

Many interesting issues remain to be explored.  Prominent among these
are the effect of long-range interactions on the transition, the 
behavior of the striped quantum Hall phase at finite temperature, 
the role of edges, the
effects of dislocations, and the effects of disorder.  These issues
are currently under study.

%In summary, we have discussed a mechanism for an
%unlocking phase transition of modulated
%sheets and stripes due to thermal and quantum fluctuations (respectively)
%in terms of a sine-Gordon model.  
%The unlocked phase is interpreted
%as a crystal phase in which unbound soliton-antisoliton pairs proliferate.
%In modulated striped phases these solitons can carry current, and
%may explain transport anomalies found recently in quantum Hall systems.

The author thanks many colleagues for useful discussions and comments,
including Luis Brey, Sankar Das Sarma, Jim Eisenstein, Charles Hanna,
and Allan MacDonald.  
The author also thanks the ITP
at UC Santa Barbara for its hospitality.  This work was supported by
NSF Grant Nos. DMR98-70681 and PHY94-07194, and by the Research
Corporation.

\begin{references}
\bibitem{dir} At present it is unclear what picks out the direction
of larger conductivity in the experiments.  In this work we assume
some weak underlying structure in the interface at which the
electron gas is located picks out a preferred direction
for the stripes.
\bibitem{dislocation} Dislocations within and between layers are
not considered in this model.  Their effects will be discussed
elsewhere.
\bibitem{com3} Hartree-Fock calculations indicate that the stripe 
modulations are quite weak, suggesting that the coupling to
disorder of the solitons is small.
\bibitem{transition} More precisely, the filling factor predicted
for the transition from the insulating
``bubble phase'' of Ref. \onlinecite{koulakov}
to the striped phase is very close to the filling factor at
which metallic behavior is first observed in Ref. \onlinecite{lilly}.
\end{references}
\end{document}